# Simulating Cloud Environments of Connected Vehicles for Anomaly Detection

M.Weiß, J. Stümpfle, F. Dettinger, N. Jazdi, M. Weyrich

University of Stuttgart

## Abstract

The emergence of connected vehicles is driven by increasing customer and regulatory demands. To meet these, more complex software applications, some of which require service-based cloud and edge backends, are developed. Due to the short lifespan of software, it becomes necessary to keep these cloud environments and their applications up to date with security updates and new features. However, as new behavior is introduced to the system, the high complexity and interdependencies between components can lead to unforeseen side effects in other system parts. As such, it becomes more challenging to recognize whether deviations to the intended system behavior are occurring, ultimately resulting in higher monitoring efforts and slower responses to errors.
To overcome this problem, a simulation of the cloud environment running in parallel to the system is proposed. This approach enables the live comparison between simulated and real cloud behavior. Therefore, a concept is developed mirroring the existing cloud system into a simulation. To collect the necessary data, an observability platform is presented, capturing telemetry and architecture information. Subsequently, a simulation environment is designed that converts the architecture into a simulation model and simulates its dynamic workload by utilizing captured communication data.
The proposed concept is evaluated in a real-world application scenario for electric vehicle charging: Vehicles can apply for an unoccupied charging station at a cloud service backend, the latter which manages all incoming requests and performs the assignment. Benchmarks are conducted by comparing the collected telemetry data with the simulated results under different loads and injected faults. The results show that regular cloud behavior is mirrored well by the simulation and that misbehavior due to fault injection is well visible, indicating that simulations are a promising data source for anomaly detection in connected vehicle cloud environments during operation.

## Introduction

Connected vehicles, i.e., vehicles able to communicate with other systems like servers or further vehicles, shape the future of transportation, enabling new use cases and functionalities which were not conceivable with the previously closed systems [1]. It is predicted that the proportion of connected vehicles in the US will exceed 85% by 2035 [2]. Thereby, the interconnection between vehicles and additions such as a cloud infrastructure open up numerous possibilities, transforming the way vehicles are perceived and interacted with. This seamless connection offers several advantages, such as real-time data exchange, enhanced vehicle performance, function offloading and improved safety measures [3].

However, the need to continuously improve the functionalities by fixing bugs or introducing new features necessitate frequent updates to ensure optimal performance and security [4]. This presents a unique set of challenges. Given the dynamics and complexity of these systems, updates could potentially lead to unforeseen anomalies, disrupting the system and resulting in unintended consequences. This is especially true in cloud backends, where services and virtualization techniques are commonly employed. While this approach offers advantages in terms of scalability and resource efficiency, it can also lead to more complex interdependencies between services [5, 6]. This problem is elevated in the connected vehicle domain by the goal to support a broad range of different functions and software variants [7]. In this environment, it becomes impossible to predict the exact software behavior and resource utilization. For example, an unexpected, large amount of user requests could lead to latencies in a single service, which, due to the system's interconnectedness, would then propagate to other services. Ultimately, such an event could convolute the whole backend.

While system metrics such as latency, CPU usage and the like can be monitored, detecting such anomalies still relies largely on expert knowledge and historical insights [8]. These however might not be accurate due to the frequent deployment of new software updates or yet unobserved changes in the environment. As such, often there is no accurate ground truth available by which anomaly detection could be performed. Ultimately, this can lead to a high monitoring effort and slower error response times, which leads to the main research question of this paper: *How can the safe and effective implementation of frequent updates in connected vehicle cloud environments be ensured?*

To overcome this challenge, the idea of this paper is to mirror the existing cloud infrastructure into a simulation. This simulated environment allows us to observe the system in parallel under normal operating conditions and assess the potential impact of updates without risking the stability of the actual system. By employing anomaly detection techniques, we can compare the simulation with the actual cloud environment, identifying any deviations that may indicate potential issues. This proactive approach enables us to address potential problems before they further manifest in the actual system, thereby ensuring the safe and efficient operation of connected vehicles.

**Contribution:** In this paper we propose an approach that enables to detect deviations from the intended behavior of a cloud environment by mirroring it into a simulation. The approach comprises:

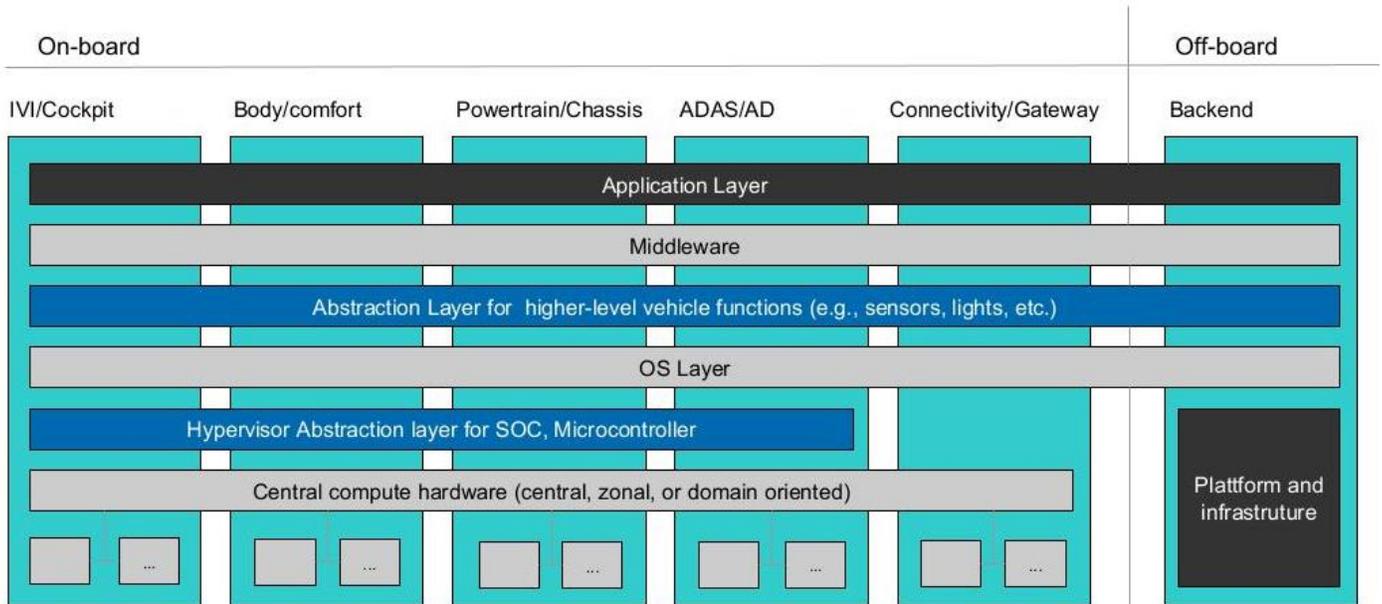
Figure 1: Zonal architecture for virtualized, connected vehicle applications [9]

- A real-world use case of the connected vehicle domain deployed to a specialized cloud
- An observability platform to continuously monitor the cloud environment during operation
- A simulation environment that automatically mirrors the cloud and enables anomaly detection

In the subsequent chapters, related work regarding connected vehicles, cloud architectures and simulators is presented. Afterwards, the cloud environment, use case, observability platform and simulation mirroring approach are described in this order. The paper closes with benchmarks for evaluation and an outlook to future work.

## Related Work

### *The Connected Vehicle Architecture of the Future*

The automotive industry stands on the precipice of a monumental shift, one that aims to realize the vision of a fully connected and autonomous vehicle. This ambitious goal necessitates a radical transition from a predominantly hardware-based industry to one that is fundamentally software-based. Central to this transformation is the concept of the Software-Defined Vehicle (SDV), a vehicle where core functionalities and features are primarily defined by software rather than hardware. This paradigm shift allows for unprecedented levels of customization, adaptability, and upgradability, effectively turning vehicles into dynamic platforms that can evolve over time [10]. The motivation behind the software-defined vehicle is manifold. Firstly, it enables rapid innovation and deployment of new features, enhancing the user experience. Secondly, it allows for real-time adaptability to changing conditions, thereby improving safety and performance. Lastly, it facilitates seamless integration with the broader Internet of Things ecosystem, unlocking new possibilities for connectivity and data-driven services.

A core enabler of the SDV is a comprehensive architecture, that not only encompasses the on-board Software and E/E-Architecture but also includes the off-board part, mainly the cloud architecture. In a concept paper the European Commission proposed a first outlook of a possible future architecture [9]. As can be seen in Figure 1, the architecture is divided into on-board and off-board. However, it is immediately noticeable that both parts are closely coupled with each other, based on the same structure. This differs greatly from current vehicles, especially with the on-board E/E-architecture. Here, a shift from a distributed or domain-centralized architecture with hundreds of ECUs towards a zonal architecture with few HPCs is seen. These HPCs are categorized into vehicle zones, such as ADAS, Cockpit, Powertrain, Body and Comfort and Connectivity. With this transition a separation between software and hardware is anticipated, acting as an enabler for the objectives of the SDV. While both, future on- and off-board architecture of the vehicle of the future are important research areas, this paper focuses on the off-board part, due to its initial objective.

At the center of cloud computing is a flexible and scalable infrastructure that provides an illusion of infinite computer resources on demand. This model is increasingly replacing the previous server-client model and enables the decoupling of flexibly combinable services. Cloud computing is closely linked to virtualized hardware, virtual data centers, software, platforms, and infrastructure as a service [11].
Virtualization makes it possible to subdivide physically fixed resources into several virtual parts. These include computing power, RAM, persistent memory and networks. In a more abstract sense, software environments such as operating systems, file systems or applications can also be virtualized.
The basis of virtualization is software, which can be divided into two main groups.

1. **Virtual Machine Monitor (VMM) software** runs on a host operating system and provides a runtime environment and hardware abstraction for guest systems. The implementation utilizes hardware allocation functions of the host operating system and is therefore considered simple. However, the diversions via these operating system functions comes with performance losses [11].
2. **Hypervisor software** as a virtualization platform, on the other hand, is not based on an operating system and implements the necessary functions itself. In the broadest sense, a virtualization platform can be seen as a highly specialized operating system. This results in reduced overheads and therefore better performance [11, 12].

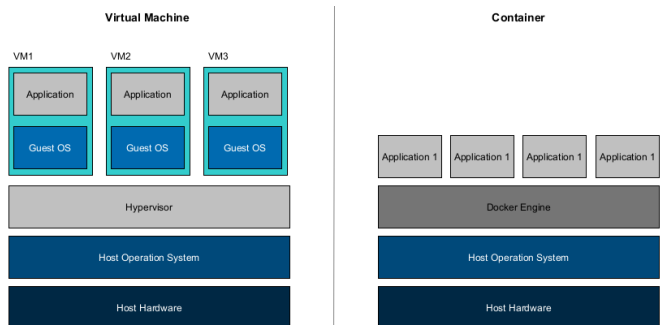

Figure 2: Comparison between Virtualization and Containerization

Next to virtualization is containerization, which differ significantly from each other. Instead of virtualizing an entire operating system and its allocated resources, a containerization layer uses the shared operating system kernel of the host operating system [13]. This difference can be seen in Figure 2.

By eliminating the hypervisor level and not having to virtualize complete operating systems, resource consumption is greatly reduced [14]. This includes, among other things, the required storage space. For example, standard Linux installations require at least several gigabytes, while the corresponding installation of such a container is in the low megabyte range. The basic resources of the virtual machines also add up, while these are largely shared between the containers and the host system in containerization. This lean concept also enables containers to be rolled out, started up and shut down quickly, which in turn increases short-term scalability. Containers remain unchangeable during operation and can therefore be replaced at any time. This also makes it possible to update the software by quickly replacing the existing container with a prepared, updated container [15].

When offering a cloud environment, different levels are proposed which are generally referred to as "X as a Service" (XaaS). These differ significantly in terms of their function and objectives [16, 17]:

- "Infrastructure as a Service" (IaaS) describes the lowest level of abstraction and provides networks, servers, data centers and storage, among other things. The physical system is managed by an IT service provider and resources are automatically and dynamically allocated to the customer. In addition to physical infrastructure, virtualized infrastructure is also frequently used.
- "Platform as a Service" (PaaS) offers tools, operating systems, databases, and middleware as a service. The actual application software comes from the customer, while the provider provides the necessary basic building blocks. This service level can therefore also be seen as a further development of classic hosting.
- "Software as a Service" (SaaS) provides standardized applications. Such offers are aimed directly at the end user. The provider takes full responsibility for the maintenance and servicing of the software.

Apart from the running applications, a key use case for containerization in a cloud environment is in the automation of the application development itself. Software is subject to a life cycle in which changes are made through the phases of development and testing right up to integration. In the context of connected vehicles, challenges arise due to the simultaneous development of several components of an application, which leads to complex branching [18]. Using container technologies, this problem can be mitigated. With continuous integration/continuous delivery (CI/CD), changes to individual parts of the software are now continuously tested by an automatic pipeline and added to the current code repository. This results in automation from code creation through to its integration into production systems [19].

### *Backend and Infrastructure Simulations*

A cloud infrastructure is typically built to meet user-specific requirements and can thus not be modified without capital and effort. Especially in research, where specific research questions are examined, the operation and modification of a real cloud infrastructure is not feasible without commercial use cases. For this reason, open-source simulators are very popular in the research community to study the communication infrastructure and backend components [20]. In this context available simulators can be divided into specific purposes e.g. energy optimization, economy, middleware, applications, and others. Available cloud simulation frameworks were analyzed in several surveys like [20–25].

GreenCloud is a packet-level simulator designed based on the network simulator NS-2 for energy-monitoring in cloud computing data centers [20, 21]. This tool offers a comprehensive model that intricately captures the energy consumption of various components within a data center, including IT equipment, computing servers, network switches and communication links. The usage of NS2 as platform leads to a small simulation reality gap but shows a poor performance with increasing size of the cloud. On the other hand, CloudNetSim ++ [22, 26] is a toolkit build on top of OMNETT++. It is used to model the energy consumption of data centers including relevant components like servers, communication links and data center infrastructure [26]. Compared to GreenCloud, CloudNetSim++ can be used to model distributed data centers.

DCSim (Data Center Simulator) is an event-driven and open-source, Java-based simulation framework designed for modeling and simulating data center management techniques and VM Placement Policy (VMPP) [27]. It is used to model continuous workloads and helps to analyze the behavior of data center components, including hosts, virtual machines, storage devices, and network elements considering specific policies.

CloudNetSim [22] is used to model end-to-end network communication between clients and server. It provides CPU scheduling and can be used to model thousands of network nodes. A Java-based simulator named CloudShed can be used to develop and evaluate multi criteria VM allocation in data centres [20].

The C++ and Java-based SimGrid VM simulator is a modified version of SimGrid [20]. It can be used to analyze large-scale distributed systems considering VM management procedures within a cloud data center and it enables the implementation and elaboration of realistic live migration techniques [20, 28]. At the same time only little hardware resources are required for extensive simulation tasks. The structure of a user's simulation environment is defined using platform files including e.g. the number of hosts and their CPU capacity.

To analyze a data centers scalability while considering performance aspects, the clouds size and design policies, the Simulation Program for Elastic Cloud Infrastructures, or in short SPECI, can be used [22]. SPECI is a discrete event simulator based on SimKit [20]. SPECIs benefit is the possibility to study inconsistencies occurred after failures appears [22].

CloudSimSDN is a Java-based framework build on CloudSim and allows the simulation of software defined network (SDN)-enabled clouds [25]. Therefor components for network traffic and SDN controllers were added to the basic CloudSim framework [29]. The focus is the implementation and evaluation of resource management policies in SDN cloud data centers and their effect on cloud infrastructure.

GroundSim is a Java-based, open-source, event-driven simulation framework designed for modeling and simulating grid and cloud computing environments [21]. It enables the analysis of the behavior of various grid and cloud entities, including resource brokers, schedulers, network elements, and user applications [30]. Additionally it is possible to model failures using background and cost models [23].

The CloudSim platform serves as a crucial generalized and extensible tool for the modeling and simulation of emerging cloud computing infrastructures and services [22, 24]. Specifically designed for the deployment of expansive cloud computing data centers. It enables the simulation of virtualized server hosts with customizable policies for resource provisioning. Moreover, the platform facilitates the simulation of application containers, contributing to a comprehensive understanding of cloud-based environments. CloudSim is a Java based platform enabling large-scale cloud simulation considering communication between cloud components. Additionally, CloudSimPlus is an extension of CloudSim and aims to simplify the simulation of data centers while increasing the accuracy of the simulation compared to CloudSim. [24]

**To summarize:** While previous papers have shown the suitability of live simulations in the analysis of vehicle software updates [31], this is not the case for vehicle cloud applications. According to Mansouri et. al [22], the most common cloud simulators in research are based on CloudSim. The other simulators presented in this chapter have not been actively used in research in recent years, so it can be concluded that they are no longer of great relevance for modern systems [25]. In addition, after analyzing various cloud simulation frameworks, the authors conclude that no framework supports mirroring of live cloud live environments, a research gap which this paper aims to close.

The following chapters are structured as follows: First, a cloud backend intended for connected vehicle functions is presented alongside with the use case of charging station lookup. Subsequently, the simulation environment is described, including an observability platform for intelligent data monitoring and a concept to mirror the cloud environment into a simulation.

# Vehicle Cloud Backend

In this chapter, the cloud backend, which was developed for this paper, is presented. First, the underlying IT infrastructure alongside with the cloud configuration is described. Afterwards, the use case application with its various services and communication paths is outlined.

*IT Infrastructure*

This section describes the underlying IT infrastructure of the implemented cloud environment. First, requirements in terms of scalability, orchestration, monitoring and management are listed. Subsequently, a suitable realization is presented.

**Requirements**

The requirements to the cloud backend IT are derived from recent advancements in the area of connected vehicles as described in the previous chapter. In total, the authors extract five key qualifications the backend should possess:

1. Due to high potential computation demands and networking, the backend should be **distributed** in nature and as such include a sufficient degree of (hardware) complexity.
2. As the number of users in a connected vehicle environment is dependent on several factors, e.g., the time of day, the backend should be able to handle a variable load of requests and as such enable a dynamic and resource-efficient **scaling** so that QoS and latency demands can be met.
3. Akin to req. 2, a flexible **communication** between cloud services, with a variable number of communication channels, should be present. Thus, client requests need to be distributed to the different backend nodes.
4. Due to the dynamic nature of the environment, the cloud backend and its applications should be highly **configurable** for flexible update options and a quick response to errors.
5. The cloud environment requires an **orchestration** layer for operations management. This includes a condensed overview of all active and inactive application parts, extensive options for redeployment and the dynamically changing any environment variables in order to respond to infrastructure changes.

**Implementation**

To fulfill the listed requirements, the backend is realized by using container technologies, allowing for easy deployment, scaling and function relocation. Figure 3 highlights the resulting architecture. In total, four different VMs are used, all of which are running on a server with an Intel Xeon E5-2630 v2 CPU. Docker is being used as **container runtime** while Kubernetes serves as underlying orchestration platform. The runtime integrates a **cluster**, consisting of multiple **nodes** responsible for task execution on hardware resources. **Worker Nodes** execute containers according to their configuration by running **kubelet** services. An **Admin Node** provides the **Control Plane**, which is responsible for the cluster management. For the latter, "Kubeadm" serves as the underlying tool, providing a Control Manager, Scheduler and a key value store ("etcd"). Communication with the admin node is established by using SSH tunneling. "Portainer" is used as the UI, providing a powerful graphical interface for managing container and cluster operations. In case of the charging use case, images are deployed as ReplicaSets, guaranteeing a baseline of available services and thus a flexible scalability. The load balancer and REST API port are set to an identical value, thus load balancing is performed on any REST API query (see next chapter).

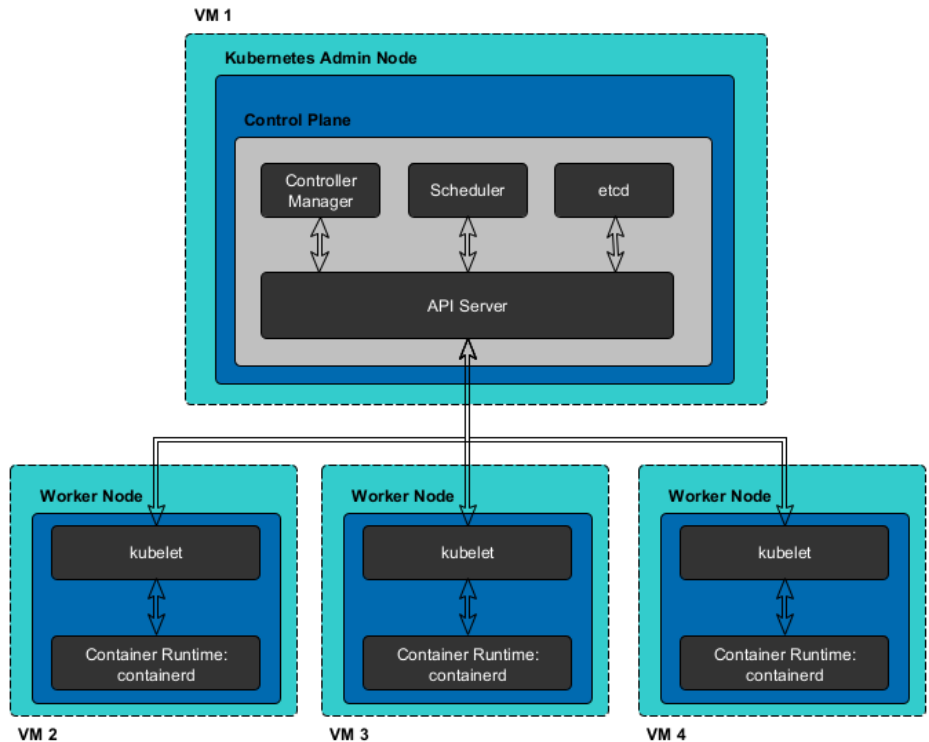
Figure 3: Cluster configuration of the Kubernetes cloud environment

## Vehicle Charging Application

To showcase the approach of this paper, a cloud application is developed which will later be mirrored by the simulation environment. In order to represent a broad range of distributed functions and communication paths for connected vehicles, several requirements are considered. First of all, a significant part of the use case must rely on a cloud environment. Some concepts for applications rely on direct communication between participants instead of cloud environments and are thus not considered. Furthermore, the use case should have a baseline complexity in terms of the variety of individual services and interfaces that make up the overall functionality. The content and purpose of the application is ultimately negligible, as it only serves to highlight the exchange between the cloud environment and its simulation. Nevertheless, a domain-specific use case is helpful to show the applicability in terms of connected vehicles.

One possible use case that fulfills the above requirements is the lookup of charging stations for electric vehicles. In this scenario, vehicles, cloud services and, ideally, all charging stations are connected. Electric charging stations offer the service of charging the battery of an electric vehicle. These charging stations have permanent properties such as their localization, number of charging points, the payment provider, the charging technology or plug offered and the respective charging capacity. Dynamic properties, that is, properties that change during runtime, are also considered and include, for example, the number of available parking spaces and - if load management is used - the currently available charging power. This data must be provided by the charging station infrastructure. The properties of an electric vehicle that are relevant in this use case include the usable charging plug, the current location, the remaining range and the currently planned route.

To facilitate an effective lookup of charging stations, it becomes necessary to consider the parameters of both the charging stations and the electric vehicles. For example, if a vehicle should be charged along its planned route, it is beneficial to check the currently available charging power of each potential charging station so that the selection can be made according to quality and efficiency demands of the user or charging infrastructure provider. In theory, one approach to implement this exchange would be to set up a fully connected network between all vehicles and all charging stations. For example, databases on the devices could be used to initially preselect suitable charging stations, which could then be queried directly for their properties. However, this has disadvantages in terms of scaling, because if many requests are sent to a device, this would slow down or completely block the system. It therefore makes more sense to set up a central, scalable backend that feeds its database with the latest data at regular intervals. This also offers the advantage that this backend knows the status of all loading points in the system and can therefore pre-process or merge data. Furthermore, it reduces the load on the computing and communication resources of the vehicles and the charging stations. For the setup of this backend the Java framework Spring Boot is chosen as a well-established solution for service-based systems.

Following the above description, an architecture can be derived which divides the functionality into a centralized charging stations service and a vehicle service, both of which are described in the following. Afterwards, the integration of the services into the cloud environment is presented.

**Charging Stations Service**

This service is intended to provide vehicles with functions relating to charging stations. To do this, first, a suitable database must be prepared. For the purpose of this use case and to obtain data from charging stations that is as close to reality as possible, a public database is used. As such, the prototype will use the publicly available charging point map of Germany's Federal Network Agency [32]. It is provided as a CSV table and, as of the day of writing, contains 54223 charging points. The following properties of charging stations are included:

- Operator
- Address (street, house number, zip code, city, federal state, district)
- Geographical information, such as latitude and longitude
- Existing charging points with plug types and charging capacity

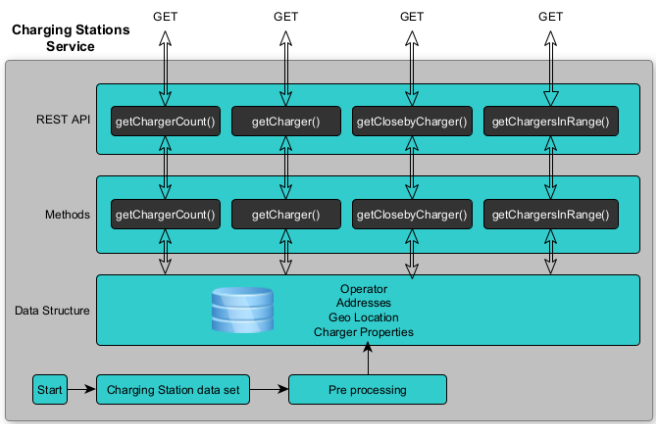

Figure 4: Charging stations service overview

In order to achieve fast data access, the table is converted into a data construct when the program is started. For this purpose, a "Charger" class is defined, which reflects all the necessary parameters of a single charging station. This class provides getter and setter functions for these parameters. At the start of program execution, the table entries are parsed and the charging station objects are created. Afterwards, the objects are kept inside a list, where the position of each object serves as a unique identifier during further runtime.

Figure 4 shows the resulting structure of the charging stations service, whose functionality is provided as a REST API. The method `getChargerCount()` returns the amount of currently available chargers by the service, whereas `getCharger()` returns the properties of an available charger with the specified unique ID. These two functions are expected to have relatively short access times, as they do not involve particularly complex algorithms and data access. More advanced functions, which accept parameters from the vehicles and require more computing power for improved service utilization, are implemented as well. The method `getClosebyCharger()` serves as a search function and receives the vehicle's location data – such as longitude and latitude – in order to find and return the nearest charging station. For this purpose, a library is used to calculate the geometric distance between the charging stations and the vehicle. Finally, the method `getChargersInRange()` extends this principle by providing a list of charging stations within the specified radius of the vehicle.

### Vehicle Service

This service is used to imitate vehicles that access the charging station service. As such, a class is implemented that allows scheduled tasks, i.e., periodic calls of the charging stations service. To accomplish this, the scheduling function of Spring Boot is used. To allow for an adaptive and dynamic environment, two parameters are defined, both of which can be adjusted in the service configuration file:

- The address of the charging station service to be accessed (relevant in scenarios with multiple cloud or charging providers)
- The call rate of methods that make API accesses.

Figure 5 shows the resulting structure of the Vehicle Service and its communication with the Charging Station Service. Based on the scheduling interval specified in the configuration, the method `getRandomCharger()` first queries the total number of charging stations in the database by the sub-method `chargerCount()`. After the result is returned, the properties of a random charger, whose ID lies in range of the charger count, are retrieved by the method `getCharger()`. In terms of advanced functionality, the method `getCloseByCharger()` first generates a random combination of geographical coordinates (limited by the area boundaries of charging station locations) and subsequently queries the backend for retrieval of a close charger.

### Cloud Integration

Both the vehicle and charging stations services are integrated into the Cloud environment as depicted in Figure 6. Load balancing is achieved by creating 3 instances of the charging station service, which run on different VMs with dedicated, isolated resources. For the purpose of this use case, an additional worker node is established on the same VM as the admin node in order to host a variable number of vehicle services that simulate dynamically changing clients.

### Simulation Environment

The depicted cloud use case is now monitored and mirrored into the simulation environment. To accomplish this, an observability platform is designed which collects the relevant data and provides it to the cloud simulator, which is extended by a module to transform the data into a simulation setup. In the following, these two parts of the simulation environment are described sequentially.

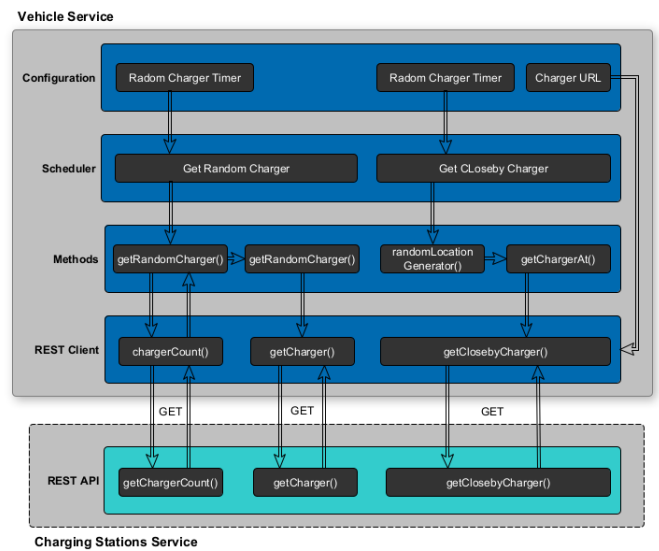

Figure 5: Vehicle service overview and interaction with the backend

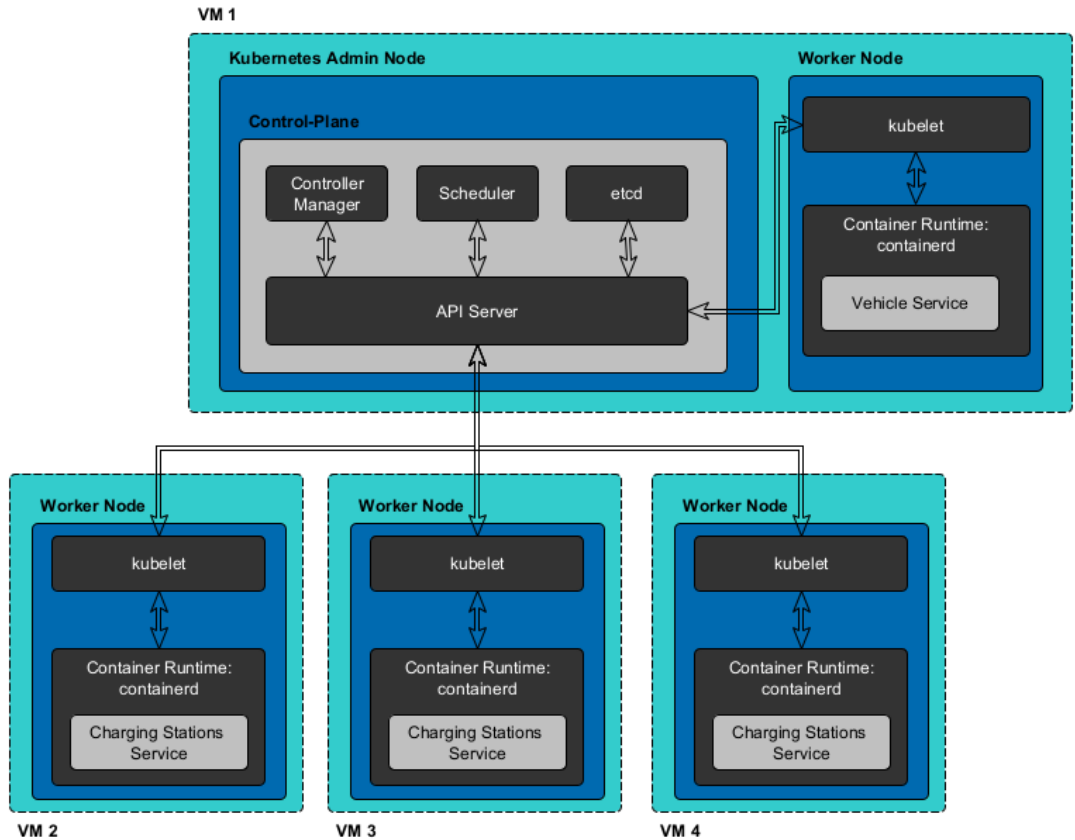

Figure 6: Integration of charging stations and vehicle services into the Kubernetes cloud environment

## Observability Platform

In order to make the data of the cloud environment visible for further analysis and at the same time support the simulation setup, extensive observability is paramount. As software solutions in this field are very specialized in certain areas and have to be adapted to the respective application, this section describes a concept for setting up an observability platform for the cloud environment and this specific use case.

The platform consists of software packages that build on each other, i.e. an "observability stack". This stack needs to collect telemetry data including logs, metrics and traces from the cloud environment, in particular the services of the use case of the vehicle and charging station services. In addition, data on the cloud environment, its structure, its applications and their distribution and resource utilization should be collected. The collected data should be stored in a suitable form and made available for retrieval. It should also be possible to visualize this data.

To realize these requirements, common practices for the creation of an observability platform are applied. As such, a Docker-Compose file in YAML format is designed, which includes all services of the observability platform, simplifying the platform deployment and allowing for its dynamic offloading if required. To facilitate independency of execution location, the platform services are isolated from the cloud environment and operate in their own Docker network. The configuration for the network and the services are kept in separate files and integrated into the services as volumes in order to create a clear separation between the container or stack configuration and the service configuration, increasing maintainability.

Figure 7 shows the implemented Observability Stack. OpenTelemetry is used for initial data retrieval of the services and the redirection to the specialized services. To accomplish this, a collector is set up, whose data pipelines consist of 3 parts: (1) Receivers for data retrieval, (2) Processors for data extraction and pre-processing and (3) Exporters for data transmission to other parts of the stack. In accordance to the requirements, data pipelines are configured for logs, metrics and traces.

After data processing by OpenTelemetry, it is transmitted towel-established tools, which are specialized in one particular data type each. Loki is used for logs, Prometheus stores metrics, Jaeger stores traces. All services are configured so that they store the respective type of telemetry and make it available for other applications via their interface. Finally, Grafana is used for accessing and visualizing the data via a graphical UI. For the purpose of this experiment, the complete Observability Stack is deployed on the Administrator Node described in previous chapters.

## Cloud Simulation

After designing the cloud environment, the use case application and the corresponding observability stack, this chapter describes the concept for mirroring the system in a simulation. First, a suitable simulation framework is selected. Subsequently, the architecture of the main application and its interaction with the cloud system are described.

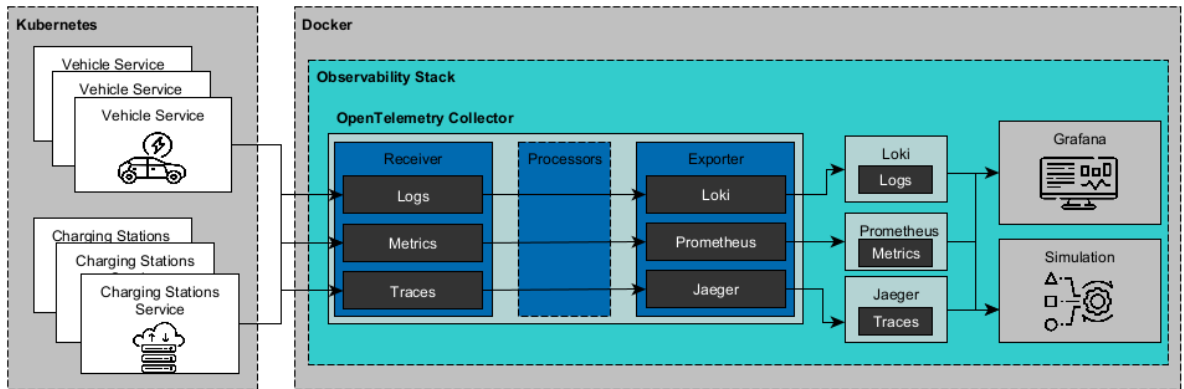

Figure 7: Observability stack for connected vehicle cloud monitoring

**Simulator Selection**

First, an existing framework for cloud systems is selected. A simulation should support VMs and applications while at the same time allow for model additions to enable possible future extensions. Furthermore, the simulation environment needs to offer extensive configurability for the automatic distribution and execution of applications on the virtual machines. The ability to execute applications at predetermined times is particularly important in order to control the timing of simulated applications based on telemetry data. The simulation environment should provide performance parameters such as CPU utilization for the individual components during and after the execution of the simulation.

As described in the related work section, most simulation environments are limited to specialized contexts such as economic efficiency, energy optimization or networking. As these do not correspond to this application scenario, they are excluded from possible candidates. Cloudlet simulators specialized in applications are more interesting, as they make it possible to determine the crucial performance parameters. However, as these are specialized for specific applications and the prototype should be kept open for all use cases, they are also not suitable. It can therefore be concluded that primarily simulation environments for general modeling are suitable. CloudSimPlus in particular should be emphasized as the de facto successor to CloudSim, offering a wide range of options for abstracting the infrastructure and architecture as well as extensive settings for the dynamic behavior of virtual machines and cloudlets, including dynamic allocation and variable time behavior. For these reasons, CloudSimPlus is used as the underlying framework in the prototype.

To enable the online simulation approach, the ability to map the architecture of the cloud environment and its applications must be created. Additionally, functions for evaluating the performance parameters and for error injection need to be provided. This functionality is implemented in a Java application that wraps the CloudSimPlus libraries. Figure 6 shows the resulting architecture and order of execution for a single simulation run, whose elements are described in the following sections.

**Architecture Retrieval**

First, a strategy on how the cloud architecture is mirrored into the simulation is required. To accomplish this, the Kubernetes API offers the option of retrieving the nodes present in the cluster together with their key properties. These properties are the number of available computing cores and working memory. A separate class is developed for accessing this information and to encapsulate the necessary parameters for establishing a connection to the Kubernetes API. Furthermore, methods are to be developed that retrieve the architecture and application data via the API.

**Retrieval of Traces**

Second, after retrieving the architecture, the executed applications are recreated. For this, two types of parameters must be retrieved: the start times of all services and the workload required to execute the application. This data can be obtained from the traces saved by the observability stack. Thus, a second class is implemented that establishes a connection to Jaeger and retrieves the corresponding telemetry.

**Simulation Setup**

After retrieval of the required data, the main simulation can be constructed. For this, the responsible class receives both the extracted architecture and the traces and instantiates a CloudSimPlus simulation. In this process, the Kubernetes architecture must be mapped to the available CloudSimPlus elements. In particular, the following elements are constructed:

- A Data Center to wrap the remaining architecture
- A Broker, which controls the dynamic behavior of the simulation
- A list of CloudSim Hosts, representing the Kubernetes Nodes
- A list of CloudSim VMs, representing the Kubernetes Pods
- A list of Cloudlets, representing services

The hosts are integrated into the data center, which in turn is made available to the broker. In addition, the lists of VMs and cloudlets are added to the broker, which can now manage them dynamically during the simulation. The list of Cloudlets is generated from the traces. As CloudSimPlus determines the workload of a cloudlet via the unit *Millions Instructions Per Second* (*MIPS*), this value must be determined based on the runtime of an application part contained in the traces and defined accordingly in the simulated cloudlet. The earliest trace recorded is defined as the starting point of the simulation. Based on this, the start time for each individual Cloudlet is calculated and specified in the simulation data structure.

After all setup steps have been performed, the simulation can finally be executed and compared against the real-world cloud data.

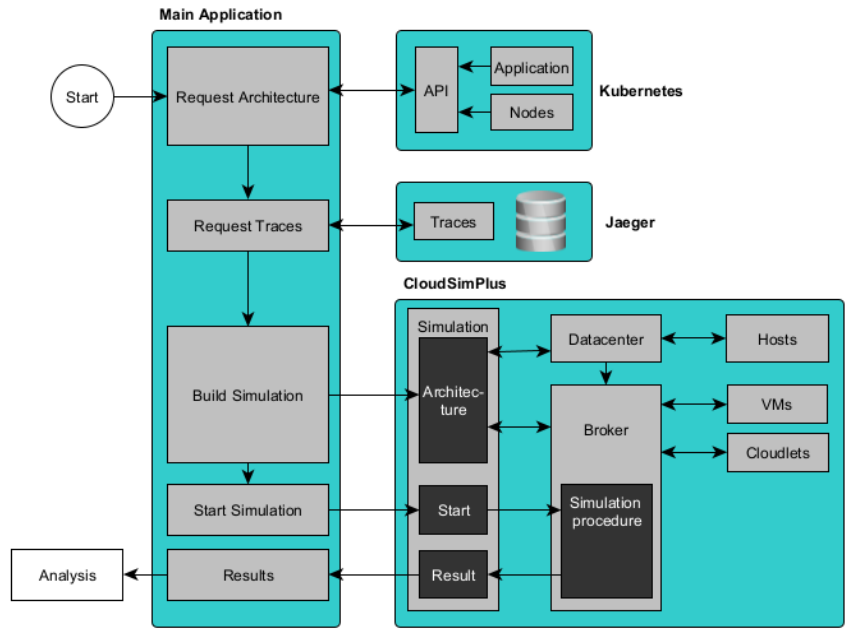

Figure 8: Cloud mirroring procedure and its interaction with CloudSimPlus

## Benchmark Results

To evaluate the implemented environment, benchmarks under variable load of the cloud system are performed. First, it is evaluated whether the simulation depicts the cloud system accurately. As such, the vehicle services are now scaled in order to apply different workloads to the charging station service and for a comparison between the respective simulation results with the real workload. For this purpose, the CPU utilization results of the simulation are compared with the real data queried by Prometheus at the respective scaling level. Figure 9 shows the simulation results across different scaling levels, displayed via the respective requests per minute. It can be inferred that the simulated CPU usage is close to the real-world data. One explanation as to why the simulation results deviate slightly downwards is that the virtual machines recorded also perform other tasks in the background in addition to the charging station services that are not covered by the simulation. These may involve communication, metrics recording or other processes, for example.

Finally, the potential of the online simulation approach for anomaly detection is evaluated. For this, a scenario is created where the cloud system behaves erroneously by injecting errors. As a result of the injection, a charging station service shuts down on the second virtual machine in the real system. The other two VMs therefore now need to take on the full load of the requests. In the meantime, the simulator still performs the intended case, i.e., all services being in operation. Figure 10 shows that the simulation continues to distribute the load across all three services on the three virtual machines as in normal operation. In the real, faulty system, however, the load on the second virtual machine is limited to the base load. The error can now successfully be detected from this data using a simple deviation analysis.

## Summary/Conclusions

In this work, a concept for simulation-based anomaly detection in the connected vehicle domain is presented. For this purpose, a use case for the retrieval of close by vehicle charging stations is implemented as a microservice architecture and containerized. This use case is deployed onto a distributed cloud environment, consisting of several virtual machines managed inside a Kubernetes cluster. This platform serves as the basis for architecture and telemetry monitoring, whose data is continuously recorded by the presented observability stack. The availability of architecture and tracing data is crucial in the strategy of mirroring the operating cloud system into the CloudSimPlus-based simulation environment.

Based on this new approach, the performance parameters of normal operation can be simulated and are thus now becoming available for anomaly detection. By the performed benchmarks it is shown that the simulation approximates the performance parameters of the cloud environment well, event under differently scaled loads. Additionally, the performed error injection leads to a clearly deviating behavior between simulation and cloud, indicating that the simulation environment can act as a suitable data source for live anomaly detection.

For the area of connected vehicles, this implies that an extended observation of the system behavior is made possible even after function extensions or updates whose effects may be difficult to trace or estimate beforehand. For this purpose, one possibility is to set up a variation of the prototype developed in this work which runs continuously in parallel to the operation of the cloud platform in order to gain real-time insights into system stability. Any misbehavior of the software system detected in this way can provide important information to system engineers to decide about necessary interventions for maintenance.

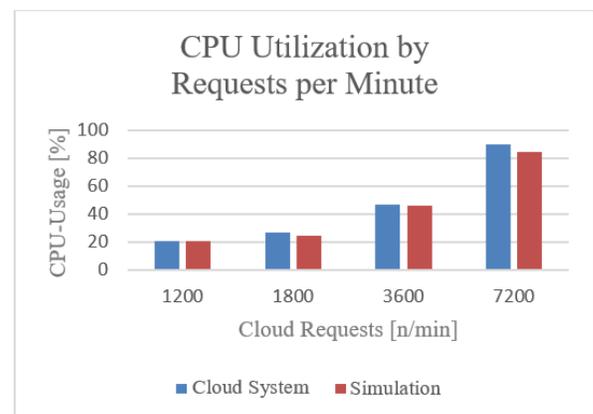

Figure 9: Comparison of CPU usage in cloud and simulator under variable load

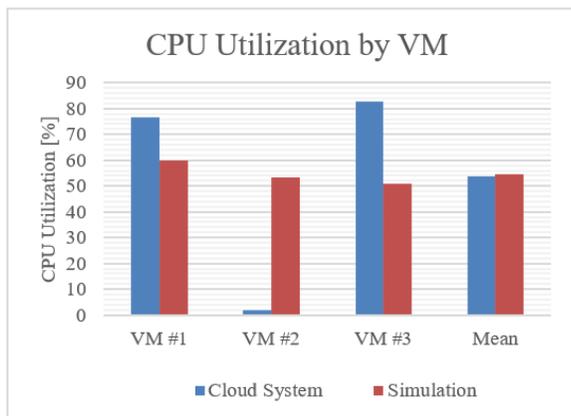

Figure 10: Comparison of CPU usage in cloud and simulator during error injection

Despite the successfulness of this approach in our simple use case however, it should still be noted that this paper should only serve as a feasibility study and does not present a sophisticated cloud simulation. The immaturity of available cloud simulators presents practical limitations to simulation results and an overall large reality-to-simulation gap that needs yet to be overcome. Additionally, the retrieval of service parameters such as MIPS can be challenging depending on the used frameworks. Thus, future work should focus on exploring how additional system KPIs can be harnessed from the simulation and how the described gap can be shrunk further, for example by integrating additional properties of the cloud environment.